\documentclass[11pt, titlepage]{amsart}
\usepackage[]{atmcs}
\usepackage{graphicx} 
\usepackage{tikz,url,hyperref,amsmath,amsthm,amssymb,pgfplots,enumitem}
\usepackage{lipsum}
\usepackage[linesnumbered,ruled,vlined,algo2e]{algorithm2e}

\def\R{\mathbb{R}}
\def\U{\mathcal{U}}
\def\res{\operatorname{res}}
\def\diam{\operatorname{diam}}
\def\norm{||}
\def\argmax{\operatorname{argmax}}
\def\mp{\operatorname{MinPts}}
\def\sub{\subseteq}
\newcommand{\s}[1]{\mathcal{#1}}

\begin{document}

\title[2-mapper]{Multiscale 2-mapper -- exploratory data analysis guided by the first Betti number}

\author{Halley Fritze}
\address{Department of Mathematics, University of Oregon\\
    Eugene, Oregon, USA}
\email{hfritze@uoregon.edu}



\begin{abstract}
    The Mapper algorithm is a fundamental tool in exploratory topological data analysis for identifying connectivity and topological clustering in data. 
    Derived from the nerve construction, Mapper graphs can contain additional information about clustering density when considering the higher-dimensional skeleta.
    To observe two-dimensional features, and capture one-dimensional topology, we construct \emph{2-Mapper}.
     A common issue in using Mapper algorithms is parameter choice.  We develop tools to choose 2-Mapper parameters that reflect persistent Betti-1 information.
     Computationally, we study how cover choice affects 2-Mapper and analyze this through a computational Multiscale Mapper algorithm.
     We test our constructions on three-dimensional shape data, including the Klein bottle.
\end{abstract}

\maketitle

\section{Introduction}

The \emph{Mapper algorithm}, constructed by Singh, et al. \cite{Mapper-OG} is a successful tool in topological data analysis used in many scientific fields including genomics \cite{genomics-mapper}, biology \cite{mapper-infection-Sasaki2020-as, covid-mapper, tda-breast-cancer}, political science \cite{mapper-mecklenburg}, machine learning \cite{mapper-nn-activations, mapper-cnn-JMLR:v24:21-0073}, and neuroscience \cite{zhang-saggar:temporalmapper, mapper-Saggar2018}. 
The Mapper algorithm is particularly useful for high-dimensional data sets, to understand clusters and relationships when
there is not enough data for training a machine
learning model.
The Mapper algorithm reflects the choice
of a filter (lens) function, with real valued functions such as coordinate functions or principal components being the most common
\cite{genomics-mapper, mapper-nn-activations, carrière2021statisticalanalysismapperstochastic, MapperParameterSelection}.
While the Mapper algorithm is originally constructed as simplicial complex derived from a nerve,
nearly all of existing mapper software \cite{KeplerMapper_v1.4.1-Zenodo, gudhi:urm,tauzin2020giottotda} only produces the one-skeleton of the nerve, called the \emph{Mapper graph}.
However, there are several applications of Mapper that use higher dimensional filters, 
\cite{tda-breast-cancer, mapper-Saggar2018,mapper-infection-Sasaki2020-as,covid-mapper, mapper-mecklenburg}, 
which opens up the possibility of incorporating
higher-dimensional geometry. 

We introduce a new Mapper object, \emph{2-Mapper} which represents the two-skeleton of the nerve of
the Mapper cover, and feature a Python implementation to display the output. 
To understand how parameter choice affects 2-Mapper, we adapt a Multiscale Mapper algorithm to compute \emph{Multiscale 2-Mapper} for data,
giving a new implementation in the standard Mapper setting as well. A basic challenge in employing Mapper is choosing parameters, since the algorithm is not stable. With Multiscale 2-Mapper, we can connect with persistent homology, and  choose scales which are representative of persistent features. 
We apply this technique to existing 3D segmentdation data \cite{3D-segmentation-data}, and provide theoretical results to the stability of Multiscale 2-Mapper given particular parameter choices.



\section{Preliminaries}



\begin{figure}[!th]
\includegraphics[width=0.8\linewidth]{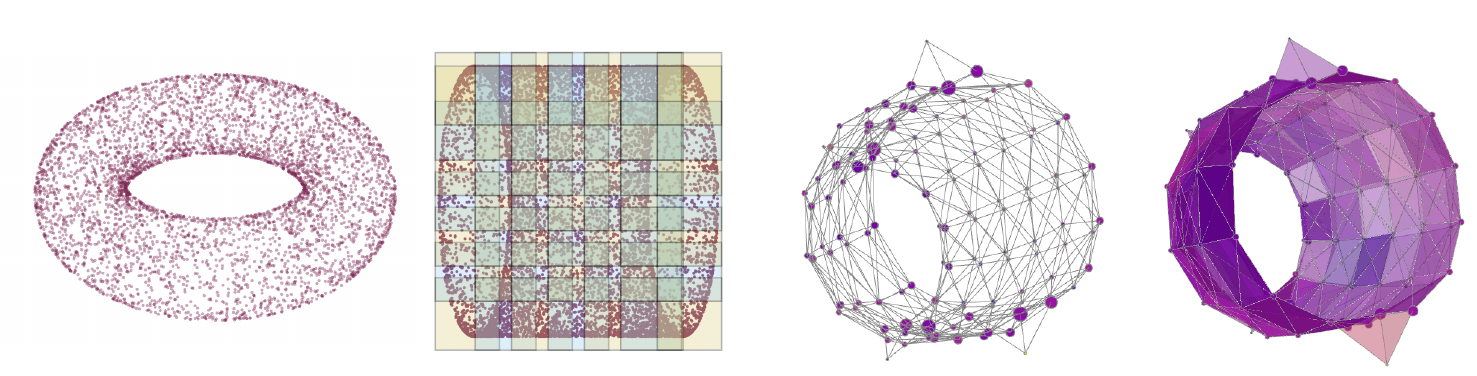}
\caption{Left to Right: A (5000 points) point cloud of the torus in $\R^3$. Compute it's Mapper graph (center right) and 2-Mapper complex (right) by projecting to $\R^2$ using coordinate projection and cubical cover with 6 intervals and 0.5 overlap fraction (middle left).}
\label{fig:2-mapper}
\end{figure}

Consider a data set $X$ with $f : X \to \R^d$ a reference function.  For some cover $\mathcal{U}$, we consider preimages of $U_i \in 
\mathcal{U}$ and cluster them.  The Mapper graph $M$ has vertices which are such clusters and edges
indicating overlap.
We introduce \emph{2-Mapper}, which incorporates data of three-fold 
overlap.
\begin{definition}[2-Mapper]
    Let $X\sub\R^d$ be a data set. 
    For choice of continuous function $f:X\to \R^m$,
    cover $\s U$ of the image $f(X)$,
    and clustering algorithm, 
    we define the \emph{2-Mapper} of $X$ as the simplicial complex
    $M(f,\s U) = \s N^2 (f^* (\s U))$.   
\end{definition}
Similar to the original Mapper construction, we can compute the first Betti number as a feature of the data \cite[Section 5.2]{Mapper-OG}.
In \figref{2-mapper} we more readily see topological features in 2-mapper and verify $\beta_0=1$ and $\beta_1=2$ for a torus point cloud.

The clustering algorithm is a key parameter. Clustering algorithms, like DBSCAN \cite{DBSCAN}, have multiple hyperparameters that lead to  variance in the Mapper algorithm.
To understand the Mapper graphs structure over a filtration of parameter choices, including both
covering and clustering parameters, we can use {\em Multiscale Mapper}, which ties the construction of Mapper to persistent homology \cite{multiscale-mappers, Topo-analysis-multiscale-mapper}. 
A filtration of cover spaces, also called a \emph{tower of covers} \cite[Definition 3]{multiscale-mappers},  gives rise to a filtration of simplicial complexes \cite[Definition 4]{multiscale-mappers}. 
We implement these ideas by  applying DBSCAN on a tower of covers $\s U$ over a data set $X$ \cite{bungula2024bifiltrationstabilitytdamapper}.
DBSCAN has hyperparameters $\epsilon$ and $\mp$, which determine search radius and minimum cluster size, respectively. Under some parameter choices, \emph{free border points} can arise while clustering a cover set $U_{\alpha,\varepsilon}\in U_\varepsilon$ \cite[Definition 8]{bungula2024bifiltrationstabilitytdamapper}.
We will choose $\mp=2$ so that no free border points when clustering $X$ with DBSCAN \cite[Corollary 1]{bungula2024bifiltrationstabilitytdamapper}.

\begin{theorem}[Tower of Cluster Covers \cite{bungula2024bifiltrationstabilitytdamapper}]\label{thm:cluster-cover}
    Let $X$ be a data set and DBSCAN be used to cluster $X$. Let $\epsilon$ and $\mp$ be fixed. If no free border points exist with respect to $\epsilon$ and $\mp$, then there is a filtration of cluster covers 
    $\{c^{U_\varepsilon,U_\delta}: C_{U_\varepsilon}\to C_{U_\delta},\ U_\varepsilon\subseteq U_\delta\}$
    where $C_{U_\varepsilon}$ is the clustered cover of $X$ with respect to the cover $U_\varepsilon$.
\end{theorem}

Using \thmref{cluster-cover}, we guarantee the existence of multiscale mapper on a data set $X$.
In section \secref{stability}, we show that cluster covers derived from $(c,s)$-good covers \cite[Definition 7]{multiscale-mappers} are also $(c,s)$-good.
Our work in \secref{multiscale-alg} constructs the multiscale mapper for default parameters in \texttt{giotto-tda} \cite{tauzin2020giottotda} and a tower of cubical covers; see \defref{tower-cubical-cover}.

\section{Multiscale Mapper and Stability for Cluster Covers}\label{sec:stability}

Consider a cover $\mathcal{U}$ over a compact topological space $Z\sub\mathbb{R}^n$, with \emph{bounding box} $\prod_{i=1}^n [m_i, M_i]$ where
$m_i = \min_{z\in Z} \pi_i(z)$ and  $M_i = \max_{z\in Z} \pi_i(z)$ for coordinate projections $\pi_i$ and $1\leq i \leq n$.
The cover we implement is a product of $1$-dimensional covers. 
 
\begin{definition}[Cubical Cover]\label{def:cubical-cover}
The \emph{cubical cover} $\mathcal{U}$ of $Z$ 
constructed with $k$-intervals and overlap fraction $g$
is the cover $\mathcal{U} = \{U_\alpha\}_{\alpha\in A}$
where
$$U_\alpha = \prod_{i=1}^n \left[c_{\alpha_i} - \frac{l_i}{2}, c_{\alpha_i} + \frac{l_i}{2}\right], 
$$
for each index $\alpha = (\alpha_1,...,\alpha_n)\in A$
where $c_{\alpha_i} = m_1 + (\alpha_i-1)(1-g)l_i + \frac{1}{2}l_i$ and $l_i = \frac{M_i-m_i}{k-(k-1)g}$
for $1\leq i\leq n$.

\end{definition}

We also call this the \emph{standard cover} since it is the only existing cover in most mapper packages in Python  \cite{KeplerMapper_v1.4.1-Zenodo, gudhi:urm, tauzin2020giottotda, mapper-interactive-wang}.  

We now extend our cover definitions to a tower of covers. This construction is straightforward, as we are extending existing cover sets by their endpoints. We can additionally think of this as a filtration of covers where we parameterize the overlap fraction $g$.

\begin{definition}[Tower of Cubical Covers]\label{def:tower-cubical-cover}
	Let $U_s = \{U_{\alpha, s}\}_{\alpha\in A}$ be a cubical cover over a compact topological space $X\in\R^n$ defined with $k$ intervals and overlap fraction $g$.
	The \emph{tower of cubical covers} $\mathcal{U}$ on $X$ 
	is a tower of covers 
	$\mathcal{U} = \{U_\varepsilon\}_{\varepsilon \geq s}$ 
	with resolution 
	$\res(\mathcal{U}) = s = \Vert{(l_1,...,l_n)}\Vert_2$.
	For $\varepsilon\geq s$ we define the cover $\mathcal{U}_\varepsilon=\{U_{\alpha,\varepsilon}\}_{\alpha\in A}$
	such that for each $\alpha\in A$
	$$ U_{\alpha,\varepsilon} = \prod_{i = 1}^n \left[c_{\alpha, i} - \frac{1}{2}\left(l_i-\varepsilon'\right), c_{\alpha, i} + \frac{1}{2}\left(l_i + \varepsilon'\right) \right],$$
	for some $\varepsilon'\geq 0$ so that 
	$\operatorname{diam}(U_{\alpha, \varepsilon}) = ||(l_1+\varepsilon',...,l_n + \varepsilon ')||_2 = \varepsilon.$
	For any $s \leq \varepsilon \leq \delta$ 
	we have canonical cover maps 
	$u_{\varepsilon, \delta}\colon U_\varepsilon \to U_{\delta}$ 
	where $U_{\alpha,\varepsilon} \to U_{\alpha, \delta}$
	for all $\alpha\in A$.
\end{definition}

Dey, M\'emoli, and Wang discuss heuristics for constructing a good tower of covers.
In particular, they remark that towers of covers constructed from expanding balls centered on a lattice are $(3,s)$-good \cite[Appendix B.2.1]{multiscale-mappers}. We provide similar results for the tower of cubical covers below for certain hyperparameters with proofs in \appendref{cover-types}.

\begin{theorem}\label{thm:cubical-good}
A tower of cubical covers $\U$ 
over a compact metric space $Z\subset\R^n$ with resolution $s$
constructed with $k$-intervals and overlap fraction $g$ such that $k\geq \sqrt{n}$
is a $(3,s)$-good tower of covers.
\end{theorem}

We extend our stability results further to verify that cluster covers that arise from using the clustering algorithm DBSCAN with a $(c,s)$-good cover is $(c,s)$-good.

\begin{theorem}\label{thm:cluster-cover-good}
	Let $\U=\{U_\varepsilon\}_{\varepsilon\geq s}$ be a tower of cubical covers with resolution $\res(\U)=s$ over a data set $X$. 
	Define the cluster cover $\s C=\{C_\varepsilon\}_{\varepsilon\geq s'}$  by clustering $X$ subordinate to the indicated cover
	using DBSCAN with fixed parameters $\epsilon$ and $\mp$.
	If there are no free border points when clustering $X$, then $\mathcal{C}$ is a $(4,s')$-good cover where $s'\leq s$.
\end{theorem}

 Achieving this stability is important, as it is rarely if ever the case that an optimal cover can be understood apriori.  
 With well-defined persistent homology, users can choose parameters for 2-Mapper complexes which exhibit persistent topological features.

\section{Multiscale Mapper implementation}\label{sec:multiscale-alg}

Like Mapper, the  2-Mapper algorithm is not stable with respect to parameter choices.  
We develop \algref{multi-mapper} to construct \emph{Multiscale 2-Mapper} to view 2-Mapper complexes over a filtration defined by parameters, in the form of a simplex tree \cite{simplex-tree} using Gudhi \cite{gudhi:urm}. 
\algref{multi-mapper} maps each cluster to its best-matched cluster in the next 2-Mapper complex in the filtration. 
Two clusters can merge as the size of the cover set increases;
to handle these cases, we insert extra edges and simplices to retain the homotopy type of the original 2-Mapper complex in the filtration, see \figref{cluster-collapse}.
In addition, because noise nodes are degenerate and not preserved across a filtration of 2-Mapper complexes we remove them from the tower of cluster covers in \thmref{cluster-cover}. 
We note that \algref{multi-mapper} can also be used to construct the original Multiscale Mapper, however with the omission of 2-simplices only the persistent $\beta_0$ barcode would contain insightful information.

We apply our algorithm to the Klein bottle data set \cite{3D-segmentation-data}. This is a data set of 15,000 points in $\R^5$. In \figref{multiscale-mapper}A, we see this data set projected in $\R^3$. As we progress through the filtration, \figref{multiscale-mapper}C, we see the inclusion of more 2-simplices in the 2-mapper complex.
For small values of $g$ ($g=0.15$), 
the 2-Mapper complex does not even resemble the shape of the Klein bottle.
However, once the overlap fraction exceeds $0.35$ the Betti-1 value is 1, and is not representative of Betti-1 for a Klein bottle.
These two observations we see are likely due to the twist in the data set.
Small overlap fractions result in smaller portions of the cover with (at least) 3-fold intersections. This leads to less edges and 2-simplices and hence a more sparse 2-Mapper complex. 
Contrarily, large overlap fractions can remove geometric nuance in the data set though the generation of too many 2-simplices. This results in the death of 1-cycles and the reduction of Betti-1.
In particular, this occurs in the twist of the Klein bottle, seen the the top left of \figref{multiscale-mapper}A. Because of the Klein bottle's self intersection, 2-Mapper constructs extra simplices in this area and this kills one of the Betti-1 representatives.
By using the persistence barcode in \figref{multiscale-mapper}B, 
we would choose $0.3\leq g\leq 0.35$ in our filtration as the best representative 2-Mapper complex
for the Klein bottle.

\section{Discussion}\label{sec:discussion}

Using our new 2-Mapper construction, we can better understand one- and two-dimensional structure  in 
data sets.
Moreover, with Multiscale 2-Mapper, we can choose
hyperparameters which capture topology at different scales.
2-Mapper and Multiscale 2-Mapper are built in Python using the \texttt{giotto-tda} and \texttt{gudhi} libraries and are currently available for use in our GitHub repository (\url{https://github.com/hfr1tz3/TwoMapper}). 
In ongoing work, we are applying 2-Mapper related to current climate data \cite{tda-weather-Strommen2023}, using the our $\beta_1$-compatible clustering as an appropriate background to analyze ``persistent'' weather patterns. The author would also like to acknowledge and thank her advisor, Dev Sinha, for his helpful insights and discussions about this work.

\pagebreak
\bibliographystyle{plain} 
\bibliography{main}

\begin{thebibliography}{10}

\bibitem{genomics-mapper}
Erik Amézquita, Farzana Nasrin, Kathleen Storey, and Masato Yoshizawa.
\newblock Genomics data analysis via spectral shape and topology.
\newblock {\em PLOS ONE}, 18:e0284820, 04 2023.

\bibitem{simplex-tree}
Jean-Daniel Boissonnat and Cl{\'e}ment Maria.
\newblock The simplex tree: An efficient data structure for general simplicial
  complexes.
\newblock {\em Algorithmica}, 70(3):406--427, Nov 2014.

\bibitem{bungula2024bifiltrationstabilitytdamapper}
Wako Bungula and Isabel Darcy.
\newblock Bi-filtration and stability of tda mapper for point cloud data, 2024.

\bibitem{carrière2021statisticalanalysismapperstochastic}
Mathieu Carrière and Bertrand Michel.
\newblock Statistical analysis of mapper for stochastic and multivariate
  filters, 2021.

\bibitem{MapperParameterSelection}
Mathieu Carrière, Bertrand Michel, and Steve Oudot.
\newblock Statistical analysis and parameter selection for mapper, 2017.

\bibitem{3D-segmentation-data}
Xiaobai Chen, Aleksey Golovinskiy, and Thomas Funkhouser.
\newblock A benchmark for 3d mesh segmentation.
\newblock {\em ACM Trans. Graph.}, 28(3), July 2009.

\bibitem{covid-mapper}
Yiran Chen and Ismar Volić.
\newblock Topological data analysis model for the spread of the coronavirus,
  Aug 2021.

\bibitem{Topo-analysis-multiscale-mapper}
Tamal~K. Dey, Facundo M{\'{e}}moli, and Yusu Wang.
\newblock Topological analysis of nerves, reeb spaces, mappers, and multiscale
  mappers.
\newblock {\em CoRR}, abs/1703.07387, 2017.

\bibitem{multiscale-mappers}
Tamal~K. Dey, Facundo Mémoli, and Yusu Wang.
\newblock {\em Multiscale Mapper: Topological Summarization via Codomain
  Covers}, pages 997--1013.

\bibitem{DBSCAN}
Martin Ester, Hans-Peter Kriegel, J\"{o}rg Sander, and Xiaowei Xu.
\newblock A density-based algorithm for discovering clusters in large spatial
  databases with noise.
\newblock In {\em Proceedings of the Second International Conference on
  Knowledge Discovery and Data Mining}, KDD'96, page 226–231. AAAI Press,
  1996.

\bibitem{mapper-mecklenburg}
Alisha Husain, Kristine Jones, Anthony Kolshorn, David Retchless, Kelemua
  Tesfaye, Courtney~M. Thatcher, and Jim Thatcher.
\newblock {\em Mappering Mecklenburg County: Exploring Census Data for
  Potential Communities of Interest}, pages 245--264.
\newblock Springer International Publishing, Cham, 2022.

\bibitem{mapper-cnn-JMLR:v24:21-0073}
Ephy~R. Love, Benjamin Filippenko, Vasileios Maroulas, and Gunnar Carlsson.
\newblock Topological convolutional layers for deep learning.
\newblock {\em Journal of Machine Learning Research}, 24(59):1--35, 2023.

\bibitem{tda-breast-cancer}
Monica Nicolau, Arnold~J Levine, and Gunnar Carlsson.
\newblock Topology based data analysis identifies a subgroup of breast cancers
  with a unique mutational profile and excellent survival.
\newblock {\em Proc. Natl. Acad. Sci. U. S. A.}, 108(17):7265--7270, April
  2011.

\bibitem{gudhi:urm}
The~GUDHI Project.
\newblock {\em GUDHI User and Reference Manual}.
\newblock GUDHI Editorial Board, 3.10.1 edition, 2024.

\bibitem{mapper-nn-activations}
Emilie Purvine, Davis Brown, Brett Jefferson, Cliff Joslyn, Brenda Praggastis,
  Archit Rathore, Madelyn Shapiro, Bei Wang, and Youjia Zhou.
\newblock Experimental observations of the topology of convolutional neural
  network activations.
\newblock {\em Proceedings of the AAAI Conference on Artificial Intelligence},
  37(8):9470--9479, Jun. 2023.

\bibitem{mapper-Saggar2018}
Manish Saggar, Olaf Sporns, Javier Gonzalez-Castillo, Peter~A. Bandettini,
  Gunnar Carlsson, Gary Glover, and Allan~L. Reiss.
\newblock Towards a new approach to reveal dynamical organization of the brain
  using topological data analysis.
\newblock {\em Nature Communications}, 9(1):1399, Apr 2018.

\bibitem{mapper-infection-Sasaki2020-as}
Karin Sasaki, Dunja Bruder, and Esteban~A Hernandez-Vargas.
\newblock Topological data analysis to model the shape of immune responses
  during co-infections.
\newblock {\em Commun. Nonlinear Sci. Numer. Simul.}, 85(105228):105228, June
  2020.

\bibitem{Mapper-OG}
Gurjeet Singh, Facundo M{\'{e}}moli, and Gunnar~E. Carlsson.
\newblock Topological methods for the analysis of high dimensional data sets
  and 3d object recognition.
\newblock In Mario Botsch, Renato Pajarola, Baoquan Chen, and Matthias Zwicker,
  editors, {\em 4th Symposium on Point Based Graphics, PBG@Eurographics 2007,
  Prague, Czech Republic, September 2-3, 2007}, pages 91--100. Eurographics
  Association, 2007.

\bibitem{tda-weather-Strommen2023}
Kristian Strommen, Matthew Chantry, Joshua Dorrington, and Nina Otter.
\newblock A topological perspective on weather regimes.
\newblock {\em Climate Dynamics}, 60(5):1415--1445, Mar 2023.

\bibitem{tauzin2020giottotda}
Guillaume Tauzin, Umberto Lupo, Lewis Tunstall, Julian~Burella Pérez, Matteo
  Caorsi, Anibal Medina-Mardones, Alberto Dassatti, and Kathryn Hess.
\newblock giotto-tda: A topological data analysis toolkit for machine learning
  and data exploration, 2020.

\bibitem{KeplerMapper_v1.4.1-Zenodo}
Hendrik~Jacob van Veen, Nathaniel Saul, David Eargle, and Sam~W. Mangham.
\newblock {\em {Kepler Mapper: A flexible Python implementation of the Mapper
  algorithm}}.
\newblock Zenodo, October 2020.

\bibitem{zhang-saggar:temporalmapper}
Mengsen Zhang, Samir Chowdhury, and Manish Saggar.
\newblock {Temporal Mapper: Transition networks in simulated and real neural
  dynamics}.
\newblock {\em Network Neuroscience}, 7(2):431--460, 06 2023.

\bibitem{mapper-interactive-wang}
Youjia Zhou, Nithin Chalapathi, Archit Rathore, Yaodong Zhao, and Bei Wang.
\newblock Mapper interactive: A scalable, extendable, and interactive toolbox
  for the visual exploration of high-dimensional data.
\newblock In {\em 2021 IEEE 14th Pacific Visualization Symposium (PacificVis)},
  pages 101--110, 2021.

\end{thebibliography}
\pagebreak
\appendix

\section{Proofs for Cover Stability}\label{append:cover-types}

\begin{lemma}\label{lem:bounding-box}
    Let $B = \prod_{i=1}^n [m_i, M_i]$ be the bounding box of $Z\subseteq\R^n$. Then $\max_{i=1,...,n} M_i - m_i \leq \diam(Z)$.
\end{lemma}
\begin{proof}
   For all $1\leq i\leq n$, $M_i-m_i$ is the length of the bounding box parallel to the $i$-th coordinate axes in $\R^n$. The diameter of $Z$ is maximum length of a line segment endpoints in $Z$. By our construction $m_i, M_i\in Z$ for all $1\leq i \leq n$, and so our lemma follows by definition.
\end{proof}


\begin{proof}[\textnormal{\textbf{\thmref{cubical-good} Proof Sketch}}]
        For condition (i), 
    and using \lemref{bounding-box}, we can show $\res(\s U) = s = \norm l \norm_2$ is such that
    $$ s \leq \frac{1}{k} \norm \vec M - \vec m\norm_2
        \leq \frac{1}{k}\norm \vec M-\vec m\norm_1 
        \leq \frac{n}{k}(M_1 - m_1) \leq \diam(Z).
    $$
    For condition (ii) let $\varepsilon\geq s$ and consider the cover $U_\varepsilon\in \U$. 
    Then for some $\varepsilon' > 0$ each cover set $U_{\alpha,\varepsilon}$
    is a hypercube with dimensions 
    $l_i^{\varepsilon'} = \frac{M_i-m_i}{k-(k-1)g} + \varepsilon'$,
     for all $\alpha\in A$.
    Choose 
    $\varepsilon' = -\frac{1}{n}\norm l\norm_1 + 
    \sqrt{ \frac{1}{n^2}\norm l\norm_1^2 
    + \frac{1}{n}(\varepsilon^2 - s^2)}$. This choice of $\varepsilon'$ is well-defined.
    For condition (iii), let $O\sub Z$ such that $\diam O> s$. 
    Let $U_{s,\alpha}\in U_s$ such that $U_{s,\alpha}\cap O \neq \emptyset$. Then there exists $x\in U_{s,\alpha} \cap O$ and center point $c_\alpha \in U_{s,\alpha}$
    \begin{align*}
    \norm c_\alpha-o\norm_2 
    \leq \norm c_\alpha - x\norm_2 + \norm x-o \norm_2 
    \leq \frac{1}{2}s + \diam(O)
    \leq \frac{3}{2}\diam(O),
    \end{align*}
    for all $o\in O$.
    Thus $O\subset U_{3\diam (O),\alpha}\in U_{3 \diam(O)}$.
    Hence $\s U$ is a $(3,\norm l\norm_2)$-good cover.
\end{proof}

\begin{proof}[\textnormal{\textbf{\thmref{cluster-cover-good} Proof}}]
	This proof follows easily from the fact that $\U$ is a $(c,s)$-good cover.
	For condition (i), note the cluster cover $C_s = \{C_{p_\alpha,s}\}$ has cover sets $C_{p_\alpha,s}\sub U_{\alpha,p}$ for all $\alpha\in A$. This means $\diam(C_{p_\alpha,s})\leq \diam(U_{\alpha,s})=s$ for all $\alpha\in A$. By definition the resolution of $\s C$ is $\res(\s C) = \sup_{\alpha_\in A} \diam(C_{p_\alpha,s})= s'\leq s$. $\U$ is $(c,s)$-good so we can conclude that $\res(\s C)\leq\res(\U)=s\leq\diam(X)$.
	Each cluster cover $C_\varepsilon=\{C_{p_\alpha,\varepsilon}\}$ for $\varepsilon\geq s$ contains cover sets $C_{p_\alpha,\varepsilon}\sub U_{\alpha,\varepsilon}$ implying that $\diam(C_{p_\alpha,\varepsilon})\leq \diam(U_{\alpha,\varepsilon})=\varepsilon$. This proves condition (ii).
	Lastly for condition (iii),
	suppose that $O\sub X$ such that $\diam(O)\geq s'$. Let $C_{p_\alpha,s}\in C_s$ such that $C_{p_\alpha,s}\cap O\neq\emptyset$. Then for $x\in C_{p_\alpha,s}\cap O$ we have for any $o\in O$ that 
	\begin{align*}
		\norm p_\alpha - o\norm_2 \leq \norm p_\alpha-x \norm_2 + \norm x-o \norm_2
		\leq s' + \diam(O)
		 \leq s + \diam(O)
		 \leq 2\diam(O)
	\end{align*}
	meaning that $O\subset C_{p_\alpha, 4\diam (O)}\in C_{4 \diam(O)}$.
	Hence $\s C$ is a $(4,s')$-good cover.
\end{proof}

\section{Algorithm for Multiscale Mapper}\label{append:multi-mapper-alg}

\begin{figure}[!h]
    \centering
    \includegraphics[width=0.8\linewidth]{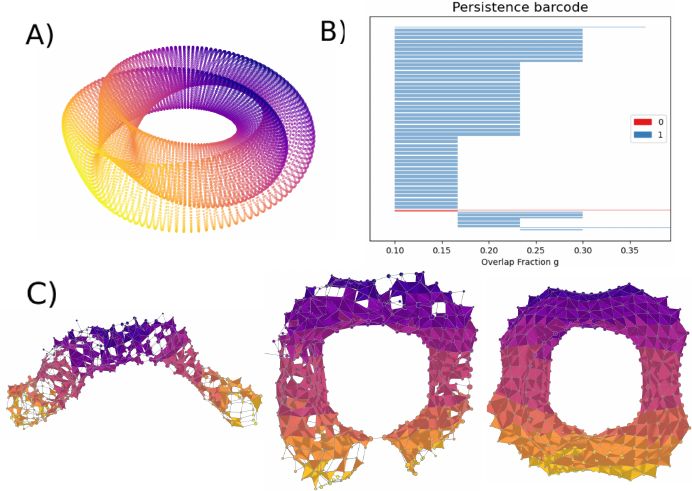}
    \caption{Multiscale 2-Mapper of the Klein bottle.
	A) The point cloud of the Klein bottle projected to $\R^3$.
	B) The persistence barcode of the multiscale 2-Mapper.
	C) 2-Mapper complexes in the filtration of the multiscale 2-Mapper. 
	From left to right are the 2-Mapper complexes with overlap fractions
	0.15, 0.2 and 0.3.}
    \label{fig:multiscale-mapper}
\end{figure}

\begin{figure}
    \centering
    \includegraphics[width=0.8\linewidth]{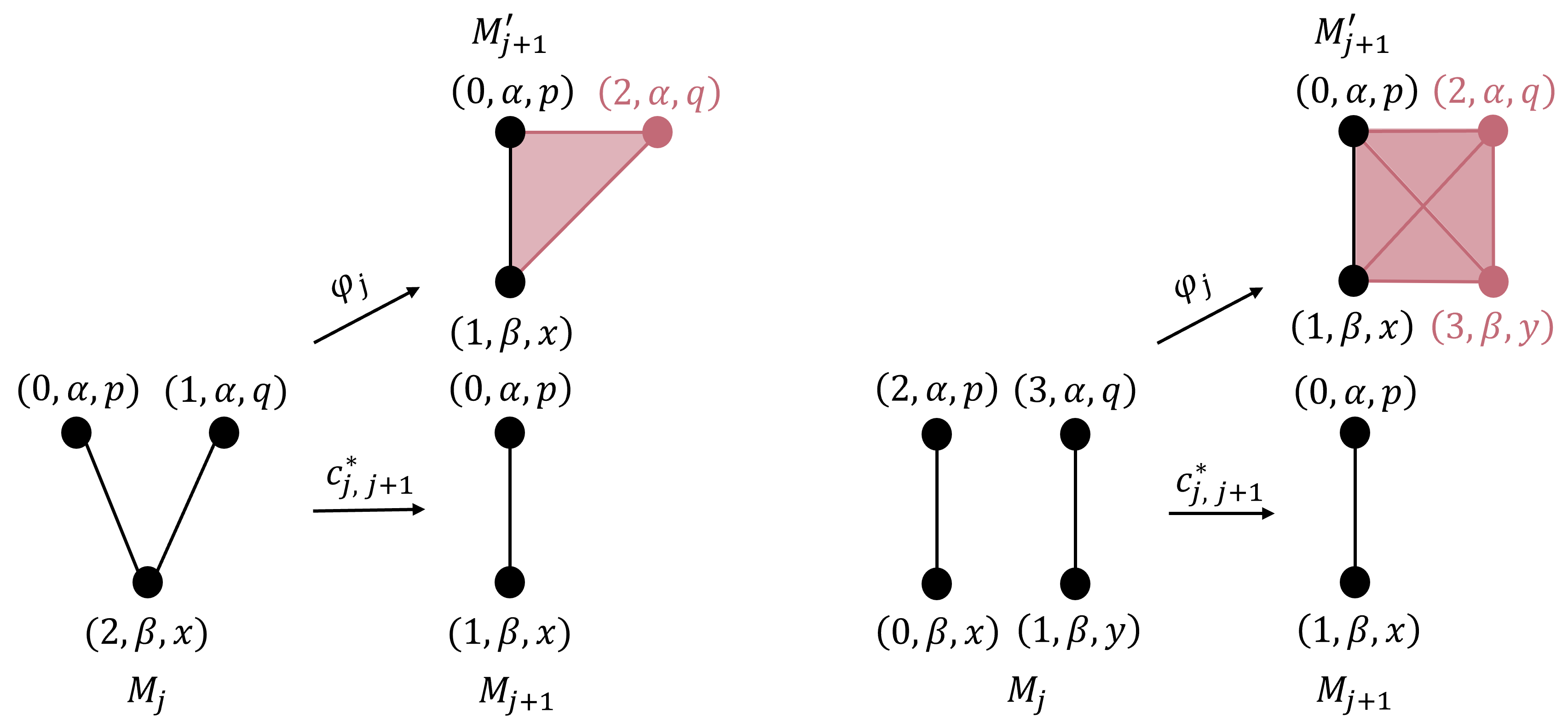}
    \caption{Cases for cluster collapse in \algref{multi-mapper}.
		Left: a cluster collapse for a single cover set.
		Right: a two cluster collapses on connected cover sets; also known as a double collapse.}
    \label{fig:cluster-collapse}
\end{figure}

Given a 2-mapper graph $M=\s N^2(f^*(U))$ on a data set $X$ with clustering algorithm DBSCAN, let $V$ be the its vertex set.
For each $\alpha\in A$, DBSCAN clusters each cover set so that $f^*(U_\alpha)=\bigcup_{i=0}^m C_{\alpha, i} \cup N_\alpha$. The classified clusters determined by DBSCAN are labeled as $C_{\alpha, i}$, and points which categorized as noise are grouped into a noise cluster $N_\alpha$.
Each node $n\in V$ is then one such cluster $C_{\alpha,i}$, and we can write $n = (X_n, \alpha, i)$ where $X_n\sub X$ is the subset of the data set in $n$.
With this notation we can define maps $\varphi_i: V_i\to V_{i+1}$ between mapper graphs $M_i$ and $M_{i+1}$ as $\varphi_i(n) = \argmax_{m\in V{i+1}} \frac{|X_n\cap X_m|}{|X_n\cup X_m|}$.
This map is surjective, however it is not necessarily injective. Injectivity fails when there are two clusters $n_1, n_2\in V_i$ that collapse into a single cluster. In this case, this means that the mapper graph $M_{i+1}$ does not contain a copy for either $n_1$ or $n_2$. Simplex trees preserve vertices across an entire filtration, and so to model this vertex collapse we instead insert the collapsed node (say $n_2$) into $M_{i+1}$, and to preserve the homotopy of $M_{i+1}$ we also insert an edges and 2-simplices depending on the adjacency of nodes with $n_2$ in the mapper graph $M_i$, see \figref{cluster-collapse}. The computational complexity for this algorithm is approximately $O((n-1)N^2)$ where $n$ is the number of 2-Mapper complexes in the Multiscale 2-Mapper and $N$ is the average number of vertices per 2-Mapper complex.

\begin{algorithm2e}[!h]
	\SetStartEndCondition{ }{}{}%
	\SetKwProg{Fn}{def}{\string:}{}
	\SetKwFunction{Range}{range}
	\SetKw{KwTo}{in}
	\SetKwFor{For}{for}{\string:}{}%
	\SetKwIF{If}{ElseIf}{Else}{if}{:}{elif}{else:}{}%
	\SetKwFor{While}{while}{:}{fintq}%
	\SetKw{Break}{break}
	\newcommand\forcond{$i$ \KwTo\Range{$n$}}
	\AlgoDontDisplayBlockMarkers\SetAlgoNoEnd\SetAlgoNoLine%
	\caption{
		Given a sequence of 2-Mapper complexes, without noise, $(M_1,...,M_n)$ computed from a tower of covers $\s U$, and clustering algorithm DBSCAN($\mp$=2, $\epsilon$) on a data set $X$, compute the Multiscale 2-Mapper. 
	}
	\label{alg:multi-mapper}
	\SetKwFunction{map}{MapperToSimplexTree}
	\SetKwFunction{align}{AlignMappers}
	\DontPrintSemicolon
	\Fn{\align($M_i$, $M_{i+1}$)}{
		\text{Let} $V_i$, $V_{i+1}$ 
		\text{be vertex sets of} $M_i$ and
		$M_{i+1}$, \text{respectively}.\;
		Let $J$ be an empty $|V_i|\times |V_{i+1}|$ matrix.\;
		\For{$(n, m)\in V_i\times V_{i+1}$}{
			Let $n = (X_n, \alpha, p_\alpha)$ and $m=(X_m, \beta, q_\beta)$.\;
			\If{$\alpha = \beta$}{
				$J_{n,m} = \frac{|X_{n}\cap X_{m}|}{|X_{n}\cup X_{m}|}$
			}
		}
		Let $\varphi_i(n) = \argmax_{m} J_{n,m}$.\;
		\For{$m\in V_{i+1}$}{
			\If{$|\varphi^{-1}_i(m)|\geq 2$  }{
				Let $n_M = \argmax_{n\in \varphi^{-1}(m)} |X_n|$.\;
				\For{$n\in \varphi^{-1}_i(m)$, \textnormal{with} $n\neq n_M$}{
					Add vertex $n\in V_{i+1}$, and set $\varphi_i(n)=n$.\;
					Add edge $(n,m) \in M_{i+1}$.\;
					\For{\textnormal{vertices} $o\in V_i$ \textnormal{adjacent to} $n$}{
						Add simplex $(o, n, m) \in M_{i+1}$.\;
					}
					\For{\textnormal{vertex pairs} $(o_1, o_2)$ \textnormal{adjacent to} $n$}{
						Add simplex $(o_1, o_2, n) \in M_{i+1}$.\;
					}
				}
				
			}
		}
		\For{\textnormal{node pairs} $(n_1,n_2)\in V_i\times V_i$}{
			\If{\textnormal{there is an edge between} $n_1$ \textnormal{and} $n_2$ 
				\textbf{\textnormal{and}} $n_1,n_2\in M_i$ \textnormal{``collapse'', ie.}
				$n_1,n_2\notin M_{i+1}$.
				}{
			Let $n'_1$ and $n'_2$ be nodes in $V_{i+1}$ that $n_1$ and $n_2$ collapse to, respectively.\;
			Add simplices $(n'_1, n'_2, n_1), (n'_1, n'_2, n_2)\in M_{i+1}$.\;
			}
		}
		\KwRet{$\varphi_i$, $M_{i+1}$}\;
	}
\end{algorithm2e}


\end{document}